\documentclass[twoside]{article}
\usepackage{fleqn,espcrc2,youngtab}

\newcommand{\AmS}{{\protect\the\textfont2
  A\kern-.1667em\lower.5ex\hbox{M}\kern-.125emS}}

\title{Stringy gravity, interacting tensionless strings and massless 
higher spins}

\author{Bo Sundborg\address{Department of Physics, Stockholm 
University \\
        Box 6730, SE-113 85 Stockholm, Sweden}%
        \thanks{E-mail: {\tt bo@physto.se}. Supported by the 
        Swedish Science Research council. Part of the talk reviews 
        work together with Parviz Haggi-Mani.}}

\begin{document}

\begin{abstract}
Consequences of a strong version of the AdS/CFT correspondence for
extremely stringy physics are examined.  In particular, properties of
\({\cal N}=4\) supersymmetric Yang-Mills theory are used to extract
results about interacting tensionless strings and massless higher spin
fields in an \( AdS_{5} \times S^5 \) background.  Furthermore, the
thermodynamics of this model signals the presence of a Hawking-Page
phase transition between \( AdS_{5} \) space and a ``black hole''-like
high temperature configuration even in the extreme string limit. 
\vspace{1pc}
\end{abstract}

\maketitle

\section{INTRODUCTION}

The properties of string theory at extreme energies has been a
deep puzzle for many years.  In the spirit of the history of
asymptotically free field theories we would expect an understanding of
short-distance symmetries to be of fundamental importance for an
efficient formulation of the theory and perhaps even for its
consistency.  Many important aspects of the problem have
been illuminated, like symmetries of the \(S\)-matrix
\cite{Gross:1988ue,Moore:1993ns} and a stringy uncertainty principle
\cite{Yoneya:1989ai,Yoneya:2000bt}, but progress has been hampered by
a fundamental obstacle: String theory is finite rather than
asymptotically free.  Finiteness means that we typically have to worry
about the full perturbation expansion, which diverges \cite{Gross:1988ib}. 
A non-perturbative formulation like field theory would then save the
day, but unfortunately a tractable field theory formulation is lacking
for closed strings.

A closed-string field theory pending, it is still true that there has
been enormous progress in the understanding of non-perturbative string
physics in recent years.  The increased knowledge has however usually
been based on dualities between low energy field theories, generalized
to the full string theories.  Only Maldacena's AdS/CFT correspondence
\cite{Maldacena:1998re,Gubser:1998bc,Witten:1998qj,Aharony:2000ti}
offers the hope of replacing an ``untractable'' string theory with a
better known quantum field theory, but it could seem unnecessarily
complicated from a high-energy perspective, because the string theory
lives in a curved AdS space.

It has been argued \cite{Isberg:1994av} that a tensionless string
theory should constitute the proper starting point for investigating
short-distance string theory, just like massless field theories are
high energy limits of massive theories.  Similarly, broken gauge
symmetries for non-zero tension could be expected to be restored in
the tensionless limit.  Quantization of tensionless strings in flat
spacetime does however prove to be problematic.  The method used in
\cite{Isberg:1994av} does not lead to a unique diagnosis, but one
possible explanation of the problems could be that the flat background
which was used is in fact inconsistent.  Evidence for this
interpretation comes from the study of theories of massless higher
spin fields
\cite{Fradkin:1987ka,Fradkin:1987ks,Fradkin:1987qy,Vasiliev:1999ba}. 
Such fields could be expected to appear in the tensionless limit of
string theory, since there are fields of arbitrarily high spin in
string theory and the masses of the free theory are all proportional
to the string tension.  Remarkably, Fradkin and Vasiliev
\cite{Fradkin:1987ks,Fradkin:1987qy} have observed that consistent
interactions of the higher spin fields require a cosmological
constant!  In fact, the theories make sense precisely in AdS
backgrounds.  Thus Maldacena's AdS/CFT conjecture actually seems to be
a very promising tool for studying the short-distance behaviour of
string theory.

The present talk addresses how the correspondence can be applied to
the study of short-distance properties of string theory.  It is mainly
based on \cite{Sundborg:2000ue,Haggi-Mani:2000ru}, which also contain
more complete references, but the emphasis is different and several
new observations are made.

\subsection{AdS/CFT}

The prime example of the AdS/CFT correspondence equates type IIB
string theory on \( AdS_{5} \times S^5 \) with a constant five-form
Ramond-Ramond field strength to \({\cal N}=4\) supersymmetric
Yang-Mills theory \cite{Maldacena:1998re}.  The rank of the Yang-Mills
gauge group \(U(N)\) is directly related to the quantized RR flux
\(N\).  For large \(N\) the Yang-Mills perturbation expansion is
conveniently organised in a topological expansion
\cite{'tHooft:1974jz} where successive terms appear with powers of
\(1/N\), and the effective loop expansion parameter within each
topology is \(g_{YM}^2 N\) rather than \(g_{YM}^2\).  For our purposes
it is also important to keep track of the length scales of the string
theory.  Modulo numerical factors the correspondence gives \(g_{YM}^2
N = R^4 T^2 \), where \(T\) is the string tension and \(R\) is the
radius of curvature of \( AdS_{5} \) and \( S^5 \).  Thus zero string
tension corresponds to a free gauge theory!  The remaining expansion
parameter is somewhat more puzzling, but can be identified with the
ratio of the Planck length to the radius of curvature, \(1/N =
\sqrt{G_{5}/R^5}\).  Note that in contrast to what one might have
expected from the conventional association of string world-sheets of
different genera to the topological expansion, \(1/N\) is \emph{not}
the ordinary string coupling (which vanishes) but a rescaled version. 
Clearly, finite \( N \) SYM provides a natural resummation of the
genus expansion!

Here it must be pointed out that although there is an impressive
amount of evidence for the Maldacena conjecture, the original argument
and most of the evidence applies to \emph{large} \(g_{YM}^2 N\), and
not to the opposite limit, which we consider here.  Still, there is
not yet any evidence against this strong form of the conjecture, and
the discussion of thermodynamics I will present towards the end 
actually justifies a belief that the the dependence on \(g_{YM}^2 N\) is 
smooth. Furthermore, we are dealing here with the most symmetric case of 
the correspondence, and many relations are protected by supersymmetry.

Nevertheless, the statement that tensionless string theory is just
\emph{free} SYM theory seems too simple.  Indeed, we will find both
actual amplitudes that can be very complicated, and a thermodynamic
behaviour which exhibits a phase transition, so something more has to
be hidden somewhere in the correspondence.  The answer to the riddle
of course lies in the way string physics is encoded in SYM theory. 
The recipes for relating generating functions
\cite{Gubser:1998bc,Witten:1998qj}, and for calculating amplitudes
\cite{Balasubramanian:1999ri,Giddings:1999qu} involve SYM correlation
functions of traces of products of fields, which generically give rise
to very complicated combinatorics, even if we only have to evaluate
free propagators.  We have a theory of \emph{interactions without
interactions}.  Dynamics intimately related to statistics and
combinatorics is what we want in a fundamental theory of gravity:
black holes as solutions to highly non-linear equations and black
holes as thermodynamic objects should appear as different sides of the
same coin.

I shall end this presentation with a description of how black hole 
thermodynamics can be reproduced, but first I want to discuss 
relevant aspects of \({\cal N}=4\) SYM at weak coupling, how 
tensionless strings appear, and how massless higher spin theory is 
embedded in SYM theory.

\section{\({\cal N}=4\) SYM AT WEAK COUPLING}

\begin{table*}[htb]
    \caption{Spectrum of relevant and marginal primaries composed of 
    scalars in terms of \(SO(6)\) Young tableaux.}\label{primaries}
\renewcommand{\tabcolsep}{1.5pc}
\renewcommand{\arraystretch}{1.5}
\begin{tabular}{@{}llll}
  \hline
Primary/Dimension& $\Delta=2$ & $\Delta=3$ & $\Delta=4$ 
% \rule[-.4cm]{0cm}{0.1cm}
\\ \hline
  $\Phi^{IJ}$ & $ {\tiny \yng(2)\oplus \bullet}$ &$ $  &$ $  
% \rule[-.4cm]{0cm}{0.1cm}
\\  
  $\Phi^{IJK}$ & $ $ & 
${\tiny\yng(3)\oplus \yng(1)\oplus\Yvcentermath1 \yng(1,1,1)}$ &$ $ 
% \rule[-.4cm]{0cm}{0.1cm}
\\ 
  $\Phi^{IJKL}$ & $ $ & $ $ & 
$ {\tiny\yng(4)\oplus 2\; \yng(2)\oplus2\bullet\oplus2\; \Yvcentermath1\yng(2,2)\oplus2\; \yng(2,1,1)\oplus\yng(1,1)}$ 
% \rule[-.4cm]{0cm}{0.1cm}
\\ 
$(\partial^{10}\Phi^{IJ})_{\mu} $ & $ $ & ${\tiny\Yvcentermath1\yng(1,1)} $ & $ $ 
% \rule[-.4cm]{0cm}{0.1cm}
\\
$(\partial^{100}\Phi^{IJK})_{\mu}$ & & $ $ & ${\tiny2\; \Yvcentermath1\yng(2,1)\oplus2\; \yng(1)}$ 
% \rule[-.4cm]{0cm}{0.1cm}
\\ 
$(\partial^{\{\Sigma n=2\}}\Phi^{IJ})_{\mu\nu}$ & $ $ & $ $ & ${\tiny\yng(2)\oplus \bullet}$ 
% \rule[-.4cm]{0cm}{0.1cm}
\\ \hline
\end{tabular}\\[2pt]
The primaries in the last row have derivatives arranged so that they 
transform as symmetric traceless Lorentz tensor, and are thus 
generalized stress tensors.
\end{table*}

In \({\cal N}=4\) SYM theory scalars, fermions and vectors are all in
the adjoint representation of the gauge group.  All local
gauge-invariant observables are polynomials in traces of products of
these adjoint fields and their derivatives.  Single-trace operators
thus constitute basic building blocks of all local observables.  In
the free theory their correlation functions may be used to compute any
other correlation functions of local observables.  Then different
fundamental fields propagate independently so it is perfectly
consistent to restrict the attention to a subset of them.  For
simplicity we only consider conformal operators built of the six
scalar fields \( \phi^{I} \) (transforming under the \(R\)-symmetry
group \(SO(6)\)).

All single-trace operators built of scalars are characterized solely 
by their index structure and how derivatives are distributed. We may 
for example write
\begin{equation}
\Phi^{IJ} \equiv {\mathrm{Tr}} \left\{ \phi^{I}\phi^{J} \right\}~,
\end{equation}
and note that the antisymmetric part of this operator vanishes by the
properties of the trace.  Derivatives inside the traces are very
significant unless they combine to equations of motion for the
fundamental fields.  In that case correlation functions vanish. 
Furthermore, overall derivatives on the operators produce new
operators, descendants in CFT parlance, whose properties can be
deduced immediately from the original operators, since they are
essentially the results of infinitesimal translations.  For instance,
the symmetric part of
\begin{equation}
\left(\partial^{10} \Phi^{IJ}\right)_{\mu} \equiv 
{\mathrm{Tr}} \left\{ \partial_{\mu} \phi^{I}\phi^{J} \right\}~,
\end{equation}
is a descendant, again by the properties of the trace.  Here we have
introduced a notation in which the symbolic exponent of the the
derivative counts how many derivatives should be applied to the
different fundamental factors in the trace.  The non-trivial operator
content of the theory is given by the primaries, which by definition
cannot be written as descendants.  The primaries can be classified in
representations of the Lorentz algebra and \( SO(6) \).  The simplest
operators are given in Table \ref{primaries}.

This table contains relevant (dimension \(\Delta = 2,~3\)) and
marginal (\(\Delta = 4\)) primaries which are important because they
may be added to the lagrangian in order to deform the theory in the
IR, without ruining the UV behaviour.  The Maldacena conjecture gives
the dimensions of all operators in the strong coupling limit from IIB
supergravity in \( AdS_{5} \times S^5 \) \cite{Witten:1998qj}.  They
agree with the dimensions of so-called chiral primaries, which are
independent of the coupling both classically and quantum-mechanically
due to non-renormalization theorems.  In our table totally symmetric
scalar primaries belong to this class.  The presence of other
operators indicates that their dimensions depend strongly on the
coupling and diverge at strong coupling.  By continuity, operators
which are relevant at zero coupling have to become marginal for some
definite intermediate value of \(g_{YM}^2 N = R^4 T^2 \), implying the
possibility of critical values of the string tension where marginal
deformations can give rise to entirely new conformal field theories
and to new realizations of the AdS/CFT correspondence.

\section{SYM AND TENSIONLESS STRINGS}

The single-trace operators admit a natural interpretation as strings
composed of string bits constituted by the fundamental fields and
their derivatives. (I use the term string bits here because it seems
to capture the situation well, but there are both differences and
similarities to Thorn's use of the term \cite{Thorn:1991fv}.) The
operators propagate to other operators, with dominant contributions
for large \( N \) corresponding to the free propagation of the
individual string-bits to other strings preserving the same cyclic
order\footnote{The actual order in one of the traces in the Green
function is reversed, since creation and annihilation operators of the
same quantum are hermitean conjugates.} of the constituent fields,
e.g.
\begin{eqnarray}
\frac{1}{N^3}
<:\!\left[{\mathrm{Tr}}(\phi^{1}\phi^{2}\phi^{3})\right]\!(x)\!:
:\!\left[{\mathrm{Tr}}(\phi^{3}\phi^{2}\phi^{1})\right]\!(0)\!:>&\nonumber\\
   =|x|^{-6}~,&
\end{eqnarray}
\begin{eqnarray}
\frac{1}{N^3}
<:\!\left[{\mathrm{Tr}}(\phi^{1}\phi^{2}\phi^{3})\right]\!(x)\!:
:\!\left[{\mathrm{Tr}}(\phi^{1}\phi^{2}\phi^{3})\right]\!(0)\!:>&&\nonumber\\
=\frac{1}{N^2}|x|^{-6}~.&&
\end{eqnarray}
This free propagation of string bits is analogous to the free 
massless propagation of points on tensionless strings \cite{Isberg:1994av}, 
although the present discussion only indirectly deals with propagation 
in the five-dimensional curved AdS space. At any rate, the shadow of 
string propagation on the four-dimensional CFT world behaves as 
expected of a tensionless string.

The single trace operators work combinatorically like holes cut out of
world-sheets in the \(1/N\) expansion\cite{'tHooft:1974jz}, and thus
correspond naturally to closed strings in correlation functions. The
cyclic property of the trace also works as a discrete remnant of the
reparametrization invariance of strings, and the effective vanishing
of Green functions of such strings with string-bits proportional to
equations of motion presumably corresponds to another remnant of the
same symmetry.

General conformal field theory correlation functions are expected to be 
expressible in terms of pairwise operator product expansions
\begin{equation}
A(x) B(y)\sim \sum_{D}
C_{AB}^{\;\;\;\;\;D} D(y)|x-y|^{\Delta_D-\Delta_A -\Delta_B}\!\!.
\label{eq:OPE}
\end{equation}
A four-point function of conformal operators can expanded in three
different channels, and for higher point functions the number of
channels increases.  In the free case one can verify explicitly that
the expressions in terms of diffferent channels agree.  As pointed out
in \cite{Haggi-Mani:2000ru}, given appropriate definitions of
amplitudes in AdS backgrounds
\cite{Balasubramanian:1999ri,Giddings:1999qu}, this duality of
conformal field theory translates to the channel duality of string
amplitudes, which was originally the defining property of string
theory \cite{Veneziano:1968yb}.  In the \(g_{YM}^2 N = R^4 T^2 = 0\)
case CFT manifestly works as a string theory.

The AdS/CFT conjecture has another, very surprising consequence when
applied to Table \ref{primaries}.  Group theory of \(SO(4,2)\) which
is the conformal group of the CFT and the isometry group of AdS,
dictates a relationship between mass and spin in AdS on one hand and
dimension and spin in the CFT on the other hand.  The bullet
(\(\bullet\)) in the last row denotes an \(SO(6)\) scalar traceless
symmetric rank 2 tensor, the scalar contribution to the stress tensor,
which is conserved.  The stress tensor corresponds to the field with
the same Lorentz transformation properties on the AdS side, i.e. the
gravitational field.  In addition, the conservation of energy-momentum
leads to the reduction of the number of degrees of freedom appropriate
for a massless field.  This is all quite natural, but the argument is
independent of \(SO(6)\) representation and therefore also applies to
the other operators in the last row of the table, which may easily be
checked to be conserved.  On the AdS side one appears to get several
charged gravitational fields!  This sounds truly strange, but only
recently have arguments been given that can rule out such spectra
\cite{Boulanger:2001rq}.  This no-go theorem however relies both on
locality and on a lagrangian formulation, assumptions which are not
fulfilled in the present case.

\section{MASSLESS HIGHER SPINS}

Generalizations of the conserved \(SO(6)\) and energy-momentum
currents
\begin{eqnarray}
\lefteqn{ {\mathrm{Tr}}\left\{ 
	\phi^I\partial_{\mu}\phi^J -\phi^J\partial_{\mu}\phi^I \right\}~,
	\label{current} }\\
\lefteqn{ {\mathrm{Tr}}\left\{ 
	(\partial_{\mu} \phi^I \partial_{\nu} \phi^I
	- {1 \over 2} \phi^I \partial_{\mu}\partial_{\nu} \phi^I)
	- {\eta_{\mu\nu} \over 4} {\rm trace}\right\} ~,}
	\label{stress}
\end{eqnarray}
with more Lorentz indices, higher spin currents, have been studied in
the literature
\cite{Berends:1984wp,Berends:1985rq,Anselmi:1999bh,Vasiliev:1999ba}. 
For example, in the condensed notation of Vasiliev odd spin currents
are
\begin{equation}
\lefteqn{ {\mathrm{Tr}}\left\{ 
	\partial_{\mu(n)}\phi^I\partial_{\mu(n+1)}\phi^J -
	(I\leftrightarrow J) \right\}~,
	\label{high-current} }
\end{equation}
where \(\mu(n)\) denotes \(n\) symmetrized indices, and the
symmetrization extends over all indices with the same label (\(\mu\)
here).  Even spin currents are instead symmetric in \(I\) and \(J\). 
There are also unique improved versions of the currents which are
traceless in the Lorentz indices \cite{Anselmi:1999bh}, as appropriate
in the present conformally symmetric case.

These currents are conserved for free fields, but not in other cases. 
Arguing in the same way as for the gravitational field we relate
masslessness and gauge symmetries in AdS to conservation laws for dual
CFT tensor currents.  For the AdS/CFT correspondence applied to
\({\cal N}=4\) SYM this means that the duals to higher spin currents
should be massless at zero tension, but not for non-zero tension.  In
fact, to each conserved higher spin current corresponds precisely one
massless higher spin gauge field in AdS \cite{Vasiliev:1999ba}.  That
masslessness occurs at zero coupling of the CFT agrees well with
expectations, since a string tension gives mass to all string states
with higher spin than two.

To form high spin operators which are conserved and correspond to
massless dual fields one needs as many derivatives as possible and few
fundamental fields.  Essentially, spin increases with number of
derivatives, and for fixed number of derivatives mass increases with
the numbers of fields.  Since we have found massless higher spins for
quadratic operators we expect all higher order operators to have
massive duals.  Due to tracelessness of the generators, linear
operators are absent for the gauge group \(SU(N)\) which has been
argued to be the correct group in the AdS/CFT correspondence
\cite{Witten:1998qj}.  Then, the full set of higher spin currents are
of the kind
\cite{Berends:1984wp,Berends:1985rq,Anselmi:1999bh,Vasiliev:1999ba}
described above.

It is interesting to compare results from the present approach based
on AdS/CFT compared to the monumental earlier achievements, mainly by
Vasiliev and co-workers.  Among many other important facts these
authors have established the structure of non-linear higher spin gauge
symmetries and found interaction terms and equations of motion
respecting the symmetries.  The structure of the higher spin fields
here is identical, including the symmetry pattern of symmetric \(IJ\)
matrices at even spin and antisymmetric at odd spin.  Thus the
proliferation of graviton fields is independent of the string origin
of our approach.  In contrast, the massive fields seem to be special
to the string theory approach.  The main and very serious drawback is
that we cannot write down a lagrangian or even the equations of motion
for the higher spin fields.  We can only calculate Green functions
which may be interpreted as a kind of AdS scattering amplitude.  On
the other hand the full perturbation expansion in terms of the
parameter \(1/N\) may be summed by just calculating Green functions
for a finite rank \(SU(N)\) gauge group.  Curiously, as we shall see
more in detail, the spectrum of massive AdS fields depends on \(N\) in
a complicated way.  It would be interesting to know if the massive
fields are more or less arbitrary additions to Vasiliev's higher spin
theories, or if they serve some purpose for consistency, perhaps at
the quantum level.  If a strong connection between the approaches is
established, the advanced mathematical machinery of higher spin gauge
theories may perhaps be applied to solve problems in large \(N\)
Yang-Mills theory.

\section{1/N COUNTING AND STRINGY BLACK HOLES}

The spectrum of a theory is conveniently encoded in its
thermodynamics, which also directly probes the physical behaviour of
the theory.  Thermodynamics plays an even more fundamental role in
theories of gravitation, because of its mysterious relation to black
holes.  In the AdS/CFT correspondence these relations have been
exploited by Witten \cite{Witten:1998qj,Witten:1998zw} who observed
that the Hawking-Page transition \cite{Hawking:1983dh} in AdS gravity
between thermal AdS space and an AdS black hole corresponds to the
deconfinement transition in Yang-Mills theory.  An essential
ingredient is that the dependence on Newton's constant of the free
energy of the black hole \(\sim R^5/G_{5} \) is translated in gauge
theory to \(N^2\) by the AdS/CFT relation \(1/N = \sqrt{G_{5}/R^5}\). 
Since the Einstein gravity argument by Hawking and Page is used,
Witten's result strictly only applies to strongly coupled gauge
theory.  The zero coupling limit of the gauge theory which will be
reviewed here was studied in \cite{Sundborg:2000ue} with the result
that the phase transition persists!  Translated back to AdS gravity,
now in the guise of tensionless string theory or higher spin theory,
some black hole like object survives even under these extreme
conditions.

First some preliminaries.  Witten's version of the AdS/CFT
correspondence relates string theory on global AdS space to gauge
theory on \(S^3\).  This makes sense if we postulate that the scalar
fields are conformally coupled to the curvature.  Then we can go
between flat space and the sphere by a conformal mapping.  Conformal
dimensions in flat space are mapped to energies on \(S^3\) (in units
of the inverse radius of curvature).  The problem of counting states
of a definite energy amounts to counting gauge invariant local
operators.  The crucial step in the counting for \(N \to \infty\)
consists in counting single trace operators or tensionless strings. 
This problem may be solved by applying a combinatorial theorem due to
P\'{o}lya, and it results in a Hagedorn-type \cite{Hagedorn:1965st}
divergence of the free energy at a critical temperature of the order
of the inverse radius.  As one may suspect the behaviour above the
Hagedorn temperature can be controlled only by taking \(N\) finite,
but then other methods are needed.

In order to count all gauge singlets, single trace or not, with the 
appropriate statistics, the idea is to first form a partition sum 
keeping track of (global) \(SU(N)\) rotations \( g \) in terms of the 
eigenvalues \( R_{i}(g) \) of the representation 
matrices \( R(g) \). Then the rotated partition sum for bosons
\begin{eqnarray}
	\lefteqn{\prod\limits_k {\prod\limits_{i=1} 
(1+x^{E_k}R_i(g)+x^{2E_k}R_i(g)^2+\ldots )} }\nonumber \\
\lefteqn{=\prod\limits_k {{\rm Det}(1-x^{E_k}R(g))^{-1}} } \nonumber \\
\lefteqn{=\exp \left( {\sum\limits_k {\mathrm{Tr} \left( {\sum\limits_{n=1}^\infty  
  {{{x^{nE_k}} \over n}R(g^n)}} \right)}} \right) }
\end{eqnarray}
can be projected to singlet states by integration over the group, using 
the orthogonality properties of group characters \( \chi(g) \). Taking also 
the fermion contribution into account yields
\begin{equation}
\int {dg}
\exp \left( 
{\sum\limits_{n=1}^\infty  
{ {\zeta_B(x^n)}-(-1)^n{\zeta_F(x^n)} \over n}\chi (g^n)}
\right)
	\label{SingletPartition}
\end{equation}
where \({\zeta_{B,F}(x^n)}\) are partition sums of the free
fundamental SYM boson and fermion fields (for \(U(1)\) gauge group).
\( \chi(g) \) is the character of the adjoint representation 
\begin{equation}
\chi (g)
=N-1+2\sum\limits_{m<n} {\cos (\alpha _m-\alpha _n)}.
	\label{AdjChar}
\end{equation}

The integrand of equation (\ref{SingletPartition}) then only depends
on a set of \(N\) eigenvalues distributed over the unit circle, and
not on their order.  By a standard large \(N\) trick such integrals
over many eigenvalues may be approximated by an integral over
eigenvalue densities \(\rho(\alpha)\).  For large \(N\) we get a
steepest descent estimate of the partition sum
\begin{equation}
Z(x)\propto \int {D\rho \,e^{-S[\rho ]}},
	\label{NPartitionFunction}
\end{equation}
by finding a \(\rho\) which makes \(S\) stationary.  In our particular
case one obtains an integral equation for \(\rho\) with qualitatively
different solutions below and above the Hagedorn temperature.  Below
the transition the stationary eigenvalue density is constant over the
whole range of eigenvalues, while above it develops ``gaps'', i.e.\
the density gets concentrated on a subset of the full range which
shrinks with increasing temperature.  Above the transition one 
finds\footnote{An almost identical calculation was done by Skagerstam 
\cite{Skagerstam:1984gv} in a different context.} for
the free energy \(F= -\log Z\) that 
\begin{eqnarray}
-{2F \over {N^2}}
\approx \zeta(x)-1+\sqrt 
{\zeta(x)^2-\zeta(x)} \nonumber &\\
 -\log \left( {\zeta(x)+\sqrt {\zeta(x)^2-\zeta(x)}} \right)&
	\label{FreeE}
\end{eqnarray}
approximately, where \(\zeta=\zeta_{B}+\zeta_{F}\).  The essential
point here is of course the \(N^2\) factor and the sharp phase
transition at the critical temperature which is similar to the
Hawking-Page transition.  Details on the behaviour close to the
transition and on the interpretation as a Hagedorn transition may be
found in \cite{Sundborg:2000ue}.

\section{CONCLUSIONS}

I have tried to demonstrate in what sense free \({\cal N}=4\)
supersymmetric Yang-Mills theory with gauge group \(SU(N)\) behaves as
an interacting theory of tensionless strings or equivalently as an
interacting theory of massless higher spin fields in an AdS
background.  In comparison to standard higher spin theory one finds
additional massive fields, whose numbers depend on the coupling.  The
construction of these theories in terms of free Yang-Mills theory 
provides evidence for their existence and finiteness, but the 
formulation is very primitive and on-shell, just like string theory. 
Hopefully, Vasiliev's \cite{Vasiliev:1999ba} more geometric formulation 
may improve the understanding, perhaps also of large \(N\) gauge theory.

We have also discussed the non-trivial physics of these models, notably 
the existence of several massless spin 2 fields, and signs of 
non-perturbative bound states analogous to black holes.


\begin{thebibliography}{9}


%\cite{Gross:1988ue}
\bibitem{Gross:1988ue}
D.~J.~Gross,
%``High-Energy Symmetries Of String Theory,''
Phys.\ Rev.\ Lett.\ {\bf 60}, 1229 (1988).
%%CITATION = PRLTA,60,1229;%%


%\cite{Moore:1993ns}
\bibitem{Moore:1993ns}
G.~Moore,
%``Symmetries of the bosonic string S matrix,''
hep-th/9310026.
%%CITATION = HEP-TH 9310026;%%


%\cite{Yoneya:1989ai}
\bibitem{Yoneya:1989ai}
T.~Yoneya,
%``On The Interpretation Of Minimal Length In String Theories,''
Mod.\ Phys.\ Lett.\ A {\bf 4}, 1587 (1989).
%%CITATION = MPLAE,A4,1587;%%


%\cite{Yoneya:2000bt}
\bibitem{Yoneya:2000bt}
T.~Yoneya,
%``String theory and space-time uncertainty principle,''
Prog.\ Theor.\ Phys.\ {\bf 103}, 1081 (2000)
[hep-th/0004074].
%%CITATION = HEP-TH 0004074;%%




%\cite{Gross:1988ib}
\bibitem{Gross:1988ib}
D.~J.~Gross and V.~Periwal,
%``String Perturbation Theory Diverges,''
Phys.\ Rev.\ Lett.\ {\bf 60}, 2105 (1988).
%%CITATION = PRLTA,60,2105;%%

%\cite{Maldacena:1998re}
\bibitem{Maldacena:1998re}
J.~Maldacena,
%``The large N limit of superconformal field theories and supergravity,''
Adv.\ Theor.\ Math.\ Phys.\ {\bf 2}, 231 (1998)
[hep-th/9711200].
%%CITATION = HEP-TH 9711200;%%



%\cite{Gubser:1998bc}
\bibitem{Gubser:1998bc}
S.~S.~Gubser, I.~R.~Klebanov and A.~M.~Polyakov,
%``Gauge theory correlators from non-critical string theory,''
Phys.\ Lett.\ B {\bf 428}, 105 (1998)
[hep-th/9802109].
%%CITATION = HEP-TH 9802109;%%



%\cite{Witten:1998qj}
\bibitem{Witten:1998qj}
E.~Witten,
%``Anti-de Sitter space and holography,''
Adv.\ Theor.\ Math.\ Phys.\ {\bf 2}, 253 (1998)
[hep-th/9802150].
%%CITATION = HEP-TH 9802150;%%



%\cite{Aharony:2000ti}
\bibitem{Aharony:2000ti}
O.~Aharony, S.~S.~Gubser, J.~Maldacena, H.~Ooguri and Y.~Oz,
%``Large N field theories, string theory and gravity,''
Phys.\ Rept.\ {\bf 323}, 183 (2000)
[hep-th/9905111].
%%CITATION = HEP-TH 9905111;%%

%\cite{Isberg:1994av}
\bibitem{Isberg:1994av}
J.~Isberg, U.~Lindstrom, B.~Sundborg and G.~Theodoridis,
%``Classical and quantized tensionless strings,''
Nucl.\ Phys.\ B {\bf 411}, 122 (1994)
[hep-th/9307108].
%%CITATION = HEP-TH 9307108;%%



%\cite{Fradkin:1987ka}
\bibitem{Fradkin:1987ka}
E.~S.~Fradkin and M.~A.~Vasiliev,
%``Candidate To The Role Of Higher Spin Symmetry,''
Annals Phys.\ {\bf 177}, 63 (1987).
%%CITATION = APNYA,177,63;%%


%\cite{Fradkin:1987ks}
\bibitem{Fradkin:1987ks}
E.~S.~Fradkin and M.~A.~Vasiliev,
%``On The Gravitational Interaction Of Massless Higher Spin Fields,''
Phys.\ Lett.\ B {\bf 189}, 89 (1987).
%%CITATION = PHLTA,B189,89;%%




%\cite{Fradkin:1987qy}
\bibitem{Fradkin:1987qy}
E.~S.~Fradkin and M.~A.~Vasiliev,
%``Cubic Interaction In Extended Theories Of Massless Higher Spin Fields,''
Nucl.\ Phys.\ B {\bf 291}, 141 (1987).
%%CITATION = NUPHA,B291,141;%%



%\cite{Vasiliev:1999ba}
\bibitem{Vasiliev:1999ba}
M.~A.~Vasiliev,
%``Higher spin gauge theories: Star-product and AdS space,''
hep-th/9910096.
%%CITATION = HEP-TH 9910096;%%


%\cite{Sundborg:2000ue}
\bibitem{Sundborg:2000ue}
B.~Sundborg,
%``The Hagedorn transition, deconfinement and N = 4 SYM theory,''
Nucl.\ Phys.\ B {\bf 573}, 349 (2000)
[hep-th/9908001].
%%CITATION = HEP-TH 9908001;%%


%\cite{Haggi-Mani:2000ru}
\bibitem{Haggi-Mani:2000ru}
P.~Haggi-Mani and B.~Sundborg,
%``Free large N supersymmetric Yang-Mills theory as a string theory,''
JHEP{\bf 0004}, 031 (2000)
[hep-th/0002189].
%%CITATION = HEP-TH 0002189;%%




%\cite{'tHooft:1974jz}
\bibitem{'tHooft:1974jz}
G.~'t Hooft,
%``A Planar Diagram Theory For Strong Interactions,''
Nucl.\ Phys.\ B {\bf 72}, 461 (1974).
%%CITATION = NUPHA,B72,461;%%






%\cite{Balasubramanian:1999ri}
\bibitem{Balasubramanian:1999ri}
V.~Balasubramanian, S.~B.~Giddings and A.~E.~Lawrence,
%``What do CFTs tell us about anti-de Sitter spacetimes?,''
JHEP{\bf 9903}, 001 (1999)
[hep-th/9902052].
%%CITATION = HEP-TH 9902052;%%



%\cite{Giddings:1999qu}
\bibitem{Giddings:1999qu}
S.~B.~Giddings,
%``The boundary S-matrix and the AdS to CFT dictionary,''
Phys.\ Rev.\ Lett.\ {\bf 83}, 2707 (1999)
[hep-th/9903048].
%%CITATION = HEP-TH 9903048;%%


%\cite{Thorn:1991fv}
\bibitem{Thorn:1991fv}
C.~B.~Thorn,
%``Reformulating string theory with the 1/N expansion,''
hep-th/9405069.
%%CITATION = HEP-TH 9405069;%%



%\cite{Veneziano:1968yb}
\bibitem{Veneziano:1968yb}
G.~Veneziano,
%``Construction Of A Crossing - Symmetric, Regge Behaved Amplitude For Linearly Rising Trajectories,''
Nuovo Cim.\ A {\bf 57}, 190 (1968).
%%CITATION = NUCIA,A57,190;%%



%\cite{Boulanger:2001rq}
\bibitem{Boulanger:2001rq}
N.~Boulanger, T.~Damour, L.~Gualtieri and M.~Henneaux,
%``Inconsistency of interacting, multigraviton theories,''
Nucl.\ Phys.\ B {\bf 597}, 127 (2001)
[hep-th/0007220].
%%CITATION = HEP-TH 0007220;%%



%\cite{Berends:1984wp}
\bibitem{Berends:1984wp}
F.~A.~Berends, G.~J.~Burgers and H.~Van Dam,
%``On Spin Three Selfinteractions,''
Z.\ Phys.\ C {\bf 24}, 247 (1984).
%%CITATION = ZEPYA,C24,247;%%




%\cite{Berends:1985rq}
\bibitem{Berends:1985rq}
F.~A.~Berends, G.~J.~Burgers and H.~van Dam,
%``On The Theoretical Problems In Constructing Interactions Involving Higher Spin Massless Particles,''
Nucl.\ Phys.\ B {\bf 260}, 295 (1985).
%%CITATION = NUPHA,B260,295;%%





%\cite{Anselmi:1999bh}
\bibitem{Anselmi:1999bh}
D.~Anselmi,
%``Theory of higher spin tensor currents and central charges,''
Nucl.\ Phys.\ B {\bf 541}, 323 (1999)
[hep-th/9808004].
%%CITATION = HEP-TH 9808004;%%


%\cite{Witten:1998zw}
\bibitem{Witten:1998zw}
E.~Witten,
%``Anti-de Sitter space, thermal phase transition, and confinement in  gauge theories,''
Adv.\ Theor.\ Math.\ Phys.\ {\bf 2}, 505 (1998)
[hep-th/9803131].
%%CITATION = HEP-TH 9803131;%%


%\cite{Hawking:1983dh}
\bibitem{Hawking:1983dh}
S.~W.~Hawking and D.~N.~Page,
%``Thermodynamics Of Black Holes In Anti-De Sitter Space,''
Commun.\ Math.\ Phys.\ {\bf 87}, 577 (1983).
%%CITATION = CMPHA,87,577;%%

%\cite{Hagedorn:1965st}
\bibitem{Hagedorn:1965st}
R.~Hagedorn,
%``Statistical Thermodynamics Of Strong Interactions At High-Energies,''
Nuovo Cim.\ Suppl.\ {\bf 3}, 147 (1965).
%%CITATION = NUCUA,3,147;%%


%\cite{Skagerstam:1984gv}
\bibitem{Skagerstam:1984gv}
B.~S.~Skagerstam,
%``On The Large N(C) Limit Of The SU(N)-C Color Quark - Gluon Partition Function,''
Z.\ Phys.\ C {\bf 24}, 97 (1984).
%%CITATION = ZEPYA,C24,97;%%




\end{thebibliography}
\end{document}